\documentstyle[12pt,a4]{article}

\newcommand{\wt}{\widetilde}
\newcommand{\ol}{\overline}

\begin{document}

\baselineskip=18pt plus 0.2pt minus 0.1pt

\begin{titlepage}
\title{\hfill\parbox{4cm}
       {\normalsize MIT-CTP-2916\\YITP-99-67\\{\tt hep-th/9910260}\\November 1999}\\
       \vspace{2cm}
       Tachyonic modes on type 0 NS5-branes%
\thanks{This work is supported in part by funds provided
by the U.S. Department of Energy (D.O.E.)
under cooperative research agreement \#DE-FC02-94ER40818
and by a Grant-in-Aid for Scientific
Research from the Ministry of Education, Science, Sports and Culture
(\#9110).}
       \vspace{1cm}}
\author{Yosuke Imamura\thanks{E-mail: \tt imamura@ctp.mit.edu}%
\\[20pt]
{\it Center for Theoretical Physics,}\\
{\it Massachusetts Institute of Technology,}\\
{\it Cambridge, MA  02139-4307, USA}
\\[7pt]
{\it and}
\\[7pt]
{\it Yukawa Institute for Theoretical Physics,}\\
{\it Kyoto University, Kyoto 606-8502, Japan}
}
\date{}

\maketitle
\thispagestyle{empty}

\vspace{0cm}

\begin{abstract}
\normalsize
Via T-duality, a stack of unwrapped type 0 NS5-branes is transformed into
a Kaluza-Klein monopole with $A_n$ type singularity at its center.
The spectrum of twisted modes at the singularity contains tachyonic modes.
We show that, in certain parameter region, this tachyonic spectrum is completely reproduced as
modes of the bulk tachyon field localized on a classical NS5-brane solution.
In passing, we show how twisted modes at the singularity reproduce
gauge fields on stacks of NS5-branes.
\end{abstract}

%\vspace{3cm}

\vfill

\noindent
%PACS codes : 11.25.Sq, 11.25.-w, 11.15.Pg.\\
%Keywords   : AdS, baryon, brane configuration, supersymmetry.

\end{titlepage}
%\tableofcontents

%%%%%%%%%%%%%%%%%%%%%%%%%%%%%%%%%%%%%%%%%%%%%%%%%%%%%%%%%%%%%%%%%%%%%%%
\section{Introduction}

NS5-branes are non-perturbative objects in string theory and,
unlike the string itself and D-branes,
we cannot directly analyze them by worldsheet CFT.
They attract great interest in connection with the mysterious interacting quantum theory
without gravity that they carry on their world volume\cite{T5,SS,MaSt,NS5holo,MS}.

The massless spectra on type IIA and type IIB NS5-branes were first obtained in \cite{CHS}.
In that work, it is shown that on an NS5-brane classical solution there are zero modes of R-R fields
representing a vector field (type IIB)
or ones representing a self-dual two-form field and a scalar field (type IIA).
These are not whole spectra on NS5-branes.
By an S-duality argument, we know that $U(Q)$ gauge symmetry is realized on type IIB NS5-branes.
Similarly, in type IIA case, it is expected that the gauge symmetry associated
with the tensor multiplet is enhanced into a `non-abelian' gauge symmetry.
For the present, however, we do not know how such a theory is described.

Type 0A and type 0B theories also contain NS5-branes.
By applying the method of \cite{CHS}, we obtain the following massless spectra.
On a type 0B NS5-brane there are
$4$ scalar fields representing fluctuations and two $U(1)$ gauge fields.
On a type 0A NS5-brane there are four fluctuation modes, one unconstrained
two-form field and two scalar fields.
If $Q$ type 0 NS5-branes coincide, in imitation of type II NS5-branes,
we can guess the gauge symmetries to be enhanced into $U(Q)\times U(Q)$.
For type 0A NS5-branes, of cause, this statement has just a superficial sense.

There is another way to obtain massless spectra on NS5-branes.
Via T-duality, unwrapped NS5-branes are transformed into Kaluza-Klein monopoles\cite{OV}.
So, we can obtain massless spectra on NS5-branes as massless modes on
Kaluza-Klein monopoles.
Recently, this method is used to obtain massless spectra
on separated type 0 NS5-branes\cite{NS5,imamura}.
In general, Kaluza-Klein monopoles have $A_n$ type singularities and,
in addition to zero modes of bulk fields,
we should take account of twisted modes of strings at the singularities.
In Section \ref{RR.sec}, we show that all `Cartan part' of gauge fields on $Q$ coincident
NS5-branes are reproduced in this way.
(For type IIA and type 0A NS5-branes, we define the `Cartan part' as the gauge symmetry
which remains when all branes separate from each other.)

As we demonstrate later, the analysis of the twisted modes shows that
tachyonic fields also exist on type 0 coincident NS5-branes if one transverse direction
is compactified on a sufficiently small circle.
The purpose of this paper is to find the counterpart of these modes in the supergravity
description of the NS5-branes.
If it is possible, they must be modes of the bulk tachyon field since
massless and massive fields cannot generate tachyonic modes.
In Section \ref{mode.sec}, we analyze modes of the bulk tachyon field on a NS5-brane classical
solution background and show that the tachyonic spectrum of twisted modes is completely reproduced,
at least in a certain parameter region.

%%%%%%%%%%%%%%%%%%%%%%%%%%%%%%%%%%%%%%%%%%%%%%%%%%%%%%%%%%%%%%%%%%%%%%%%%%%
\section{T-duality of NS5-branes}
In this section, we briefly review the T-duality between unwrapped NS5-branes
and Kaluza-Klein monopoles with keeping our eyes on geometry and symmetry.
First, we discuss familiar T-duality, namely,
0A/0B and IIA/IIB duality of
unwrapped NS5-branes in ${\bf S}^1$
compactified string theory.
We will mention the `exotic' IIA/0B and IIB/0A duality suggested in \cite{BG}
at the end of this section.

The following arguments applicable to type IIA, IIB, 0A and 0B theories.
Let us assume NS5-branes extending along $x^0,\ldots,x^5$
and the $x^9$ direction compactified on ${\bf S}^1$ with radius $R$.
These NS5-branes are magnetically coupled to NS-NS $2$-form field $B_{i9}$ ($i=6,7,8$).
Because $B_{i9}$ is transformed into the metric component $g_{i9}$ by T-duality,
the dual to a stack of $Q$ NS5-branes is a Kaluza-Klein monopole with charge $Q$\cite{OV}.

The metric of generic parallel Kaluza-Klein monopoles is given by\cite{Hawking}
\begin{equation}
ds^2=\eta_{\mu\nu}dx^\mu dx^\nu+v(x^i)(dx^i)^2+v^{-1}(x^i)(dx^9+A_i(x^i)dx^i)^2,
\label{KKM}
\end{equation}
where $x^\mu$ ($\mu=0,\ldots,5$) and $x^i$ ($i=6,7,8$) are coordinates
of the parallel directions and the transverse directions, respectively.
The coordinate $x^9$ parameterizes the compactified direction and takes value in
$0\leq x^9<2\pi\wt R$, where $\wt R$ is compactification radius.
If we denote the monopole density by $\rho_{\rm mon}(x^i)$,
a magnetic potential $v(x^i)$ and an electric potential $A_i(x^i)\sim g_{9i}$
are obtained by solving the following differential equations.
\begin{equation}
\sum_{i=6}^8\left(\frac{\partial}{\partial x^i}\right)v(x^i)=-2\pi\wt R\rho_{\rm mon}(x^i),\quad
\epsilon_{ijk}\partial_jA_k(x^i)=\partial_iv(x^i).
\end{equation}
When the charge $Q$ concentrates at $x^i=0$,
by setting $\rho_{\rm mon}(x^i)=Q\delta^3(x^i)$, we obtain
\begin{equation}
v(x^i)=1+\frac{\wt RQ}{2|x^i|}.
\label{vis}
\end{equation}
In a part of this manifold where $|x^i|$ is much smaller than $\wt RQ$,
we can neglect the first term on the left hand side of (\ref{vis}).
If we introduce a radial coordinate $r=\sqrt{2\wt RQ|x^i|}$ representing the geodesic distance from
the point $x^i=0$, and the line element $d\Omega_2$ on unit ${\bf S}^2$,
(\ref{KKM}) is rewritten as
\begin{equation}
ds^2=\eta_{\mu\nu}dx^\mu dx^\nu+dr^2
+\frac{r^2}{4}\left[d\Omega_2^2
             +\frac{4}{\wt R^2}\left(\frac{dx^9}{Q}+a\cdot d\Omega_2\right)^2\right],
\label{flat}
\end{equation}
where $a_i$ is the unit charge magnetic monopole gauge configuration on the unit ${\bf S}^2$
and $A_i$ is written as $A_i=(Q/|x^i|)a_i$.
The metric (\ref{flat}) represents an orbifold ${\bf R}^4/{\bf Z}_Q$ since
if the period of $x^9$ were $0\leq x^9<2\pi\wt RQ$, which is $Q$ times what it actually is,
it would represent flat ${\bf R}^4$\cite{gh}.
By moving to a Cartesian coordinate, we can show that the ${\bf Z}_Q$ is subgroup
of one of two $SU(2)$ factors of $SO(4)$ symmetry of the divided ${\bf R}^4$.
We will refer to the $SU(2)$ containing ${\bf Z}_Q$ as $SU(2)_L$ and
to the other as $SU(2)_R$.
If $Q\geq3$, $SU(2)_L$ is broken to $U(1)$ by the ${\bf Z}_Q$ orbifolding.
Let $U(1)_L$ denote this symmetry.
Even if $Q\leq2$, $SU(2)_L$ is broken to $U(1)_L$ on the whole manifold described
by the metric (\ref{KKM}).
This $U(1)_L$ symmetry represents a shift of the $x^9$ coordinate.
On the other hand, $SU(2)_R$ represents rotation of the base ${\bf R}^3$ parameterized by $x^i$.
Therefore, the isometry of the Kaluza-Klein monopole solution is
\begin{equation}
P(1,5)\times SU(2)_R\times U(1),
\label{KKMsym}
\end{equation}
where $P(1,5)$ represents the Poincar\'e symmetry on a $1+5$-dimentional Minkowski space.

The symmetry of NS5-branes in uncompactified ten-dimensional spacetime
is $P(1,5)\times SO(4)$.
Let $SU(2)_1$ and $SU(2)_2$ denote the two $SU(2)$ factors of this $SO(4)$.
Due to compactification of one transverse direction,
$SO(4)$ is broken to $SO(3)\sim SU(2)$.
We call it $SU(2)_D$ since it is the diagonal subgroup of $SU(2)_1\times SU(2)_2$.
Therefore, the symmetry of a NS5-brane configuration in ${\bf S}^1$ compactified spacetime is
\begin{equation}
P(1,5)\times SU(2)_D.
\end{equation}
In the low-lying supergravity level, it is known that
${\bf S}^1$ compactified type IIA theory and type IIB theory
have the same effective action\cite{BHO}.
Because the classical Kaluza-Klein monopole solution (\ref{KKM}) does not depend on $x^9$,
it is also a solution of nine dimensional supergravity.
Therefore, by the T-duality of supergravity,
we can obtain a solution in nine-dimensional supergravity
coupled to $B_{i9}$ magnetically.
This solution is lifted up to ten-dimensional solution independent from $x^9$.
It is called a `smeared' NS5-brane solution.
The symmetry of this solution is
\begin{equation}
P(1,5)\times SU(2)_D\times U(1)_S,
\label{PDS}
\end{equation}
where $U(1)_S$ represent a shift along $x^9$ direction.

Let us compare the symmetry of the NS5-brane configuration and the Kaluza-Klein monopole
configuration.
The symmetry $P(1,5)\times SU(2)_D$ of the NS5-branes is
identified with the symmetry $P(1,5)\times SU(2)_R$ on the Kaluza-Klein monopole side.
Because $U(1)_L$ and $U(1)_S$ factors in (\ref{KKMsym}) and (\ref{PDS})
are associated with the gauge field $g_{9i}$,
they are transformed by T-duality into non-geometric symmetries
associated with $B_{i9}$.

Finally, we will mention the exotic T-dualities.
To relate type 0 theory and type II theory by the T-duality,
we should introduce non-trivial monodromies around ${\bf S}^1$.
Namely, we need monodromy $(-1)^{F_R}$ for type 0 theory and
$(-1)^{F_S}$ for type II theory,
where the operators $F_R$ and $F_S$ are the right moving worldsheet fermion number
and the spacetime fermion number, respectively\cite{BG}.
In what follows, we call a circle with monodromy $(-1)^{F_S}$ as ${\bf S}^1_{(F_S)}$
and one with monodromy $(-1)^{F_R}$ as ${\bf S}^1_{(F_R)}$.
Via this duality, a stack of $Q$ unwrapped NS5-branes is transformed
into a Kaluza-Klein monopole with charge $2Q$, rather than $Q$\cite{imamura}.
The central part of the Kaluza-Klein monopole
is described as an orbifold ${\bf R}^4/{\bf Z}_{2Q}$.
Because of the non-trivial monodromy of the ${\bf S}^1$ cycle,
the dividing ${\bf Z}_{2Q}$ is generated by $(-1)^{F_S}\gamma$ (type II) or
$(-1)^{F_R}\gamma$ (type 0), where $\gamma$ is a generator of
${\bf Z}_{2Q}\subset SU(2)_L$.
In what follows, we call them
${\bf Z}^{(F_S)}_{2Q}$ (generated by $(-1)^{F_S}\gamma$)
and ${\bf Z}^{(F_R)}_{2Q}$ (generated by $(-1)^{F_R}\gamma$).
These duality relations between NS5-branes and Kaluza-Klein monopoles
are summarized in Table \ref{dual.tbl}.
\begin{table}[htb]
\caption{T-duality between unwrapped NS5-branes and Kaluza-Klein monopoles.}
\label{dual.tbl}
\begin{center}
\begin{tabular}{cc}
\hline\hline
$Q$ unwrapped NS5-branes & Kaluza-Klein monopoles \\
\hline
\\[-2ex]
type IIA on ${\bf S}^1$ & type IIB on ${\bf R}^4/{\bf Z}_Q$ \\
type IIB on ${\bf S}^1$ & type IIA on ${\bf R}^4/{\bf Z}_Q$ \\[1ex]
type IIA on ${\bf S}^1_{(F_S)}$ & type 0B on ${\bf R}^4/{\bf Z}^{(F_R)}_{2Q}$ \\
type IIB on ${\bf S}^1_{(F_S)}$ & type 0A on ${\bf R}^4/{\bf Z}^{(F_R)}_{2Q}$ \\[1ex]
type 0A on ${\bf S}^1$ & type 0B on ${\bf R}^4/{\bf Z}_Q$ \\
type 0B on ${\bf S}^1$ & type 0A on ${\bf R}^4/{\bf Z}_Q$ \\[1ex]
type 0A on ${\bf S}^1_{(F_R)}$ & type IIB on ${\bf R}^4/{\bf Z}^{(F_S)}_{2Q}$ \\
type 0B on ${\bf S}^1_{(F_R)}$ & type IIA on ${\bf R}^4/{\bf Z}^{(F_S)}_{2Q}$ \\[1ex]
\hline
\end{tabular}
\end{center}
\end{table}

%%%%%%%%%%%%%%%%%%%%%%%%%%%%%%%%%%%%%%%%%%%%%%%%%%%%%%%%%%%%%%%%%%%%%%%%%%%
\section{Tachyonic twisted modes}\label{NSNS.sec}
In this section, we give a tachyonic part of the spectrum
of twisted modes of orbifold ${\bf R}^4/\Gamma$,
where $\Gamma$ is one of ${\bf Z}_N$, ${\bf Z}^{(F_S)}_N$ and ${\bf Z}^{(F_R)}_N$.
The integer $N$ is equal to $Q$ or $2Q$ depending on the type of T-duality.
We need to consider only the NS-NS sector since other sectors does not contain
tachyonic modes.

Let us introduce complex coordinates $z^A$ ($A=1,2$) on the divided ${\bf R}^4$,
which is rotated by $SU(2)_L\times SU(2)_R$ as
\begin{equation}
U\rightarrow g_LUg_R,\quad
U\equiv
\left(\begin{array}{cc}
z^1 & z^2 \\
-\ol z_2 & \ol z_1
\end{array}\right),\quad
g_L\in SU(2)_L,\quad
g_R\in SU(2)_R.
\end{equation}
A generator $\gamma$ of discrete group $\Gamma$ acts on the complex coordinates as follows.
\begin{equation}
\gamma :\quad
z^A\rightarrow\exp\left(\frac{2\pi i}{N}\right)z^A,\quad
\ol z_A\rightarrow\exp\left(-\frac{2\pi i}{N}\right)\ol z_A.
\end{equation}
At the fixed point of this orbifold, $N-1$ twisted modes appear.
We label them as $k=1,\ldots,N-1$
For each $k$, $z^A$ and $\ol z_A$ satisfy the boundary conditions
\begin{equation}
z^A(\sigma+2\pi)=e^{2\pi ik/N}z^A(\sigma),\quad
\ol z_A(\sigma+2\pi)=e^{-2\pi ik/N}\ol z_A(\sigma),
\label{bosonbc}
\end{equation}
and are expanded as
\begin{eqnarray}
z^A&=&l_s\sum_{m=-\infty}^\infty\left(\frac{ie^{i(m-k/N)(\tau-\sigma)}}{m-k/N}z^A_{m-k/N}
  +\frac{ie^{i(m+k/N)(\tau+\sigma)}}{m+k/N}\wt z^A_{m+k/N}\right),\label{Zexp}\\
\ol z_A&=&l_s\sum_{m=-\infty}^\infty\left(\frac{ie^{i(m+k/N)(\tau-\sigma)}}{m+k/N}\ol z_{A,m+k/N}
  +\frac{ie^{i(m-k/N)(\tau+\sigma)}}{m-k/N}\ol{\wt z}_{A,m-k/N}\right).\label{ZBexp}
\end{eqnarray}
In a same way, boundary conditions and expansions of NSR fermions $\psi^A$ and $\ol\psi_A$,
which are superpartners of $z^A$ and $\ol z_A$, are given by
\begin{equation}
\psi^A(\sigma+2\pi)=-e^{2\pi ik/N}\psi^A(\sigma),\quad
\ol\psi_A(\sigma+2\pi)=-e^{-2\pi ik/N}\ol\psi_A(\sigma),
\end{equation}
and
\begin{eqnarray}
\psi^A&=&\sum_{m=-\infty}^\infty
\left(\psi^A_{m+\frac{1}{2}-\frac{k}{N}}e^{i(m+\frac{1}{2}-\frac{k}{N})(\tau-\sigma)}
+\wt \psi^A_{m+\frac{1}{2}+\frac{k}{N}}e^{i(m+\frac{1}{2}+\frac{k}{N})(\tau+\sigma)}\right),\\
\ol\psi_A&=&\sum_{m=-\infty}^\infty
\left(\ol\psi_{A,m+\frac{1}{2}+\frac{k}{N}}e^{i(m+\frac{1}{2}+\frac{k}{N})(\tau-\sigma)}
+\ol{\wt\psi}_{A,m+\frac{1}{2}-\frac{k}{N}}e^{i(m+\frac{1}{2}-\frac{k}{N})(\tau+\sigma)}\right).
\end{eqnarray}
First, we should give the energy and the $U(1)_L$ charge of ground state.
$U(1)_L$ charge is necessary when we carry out the $\Gamma$ projection.
For this purpose, we can use the following $\zeta$-function regularization formulae.
\begin{equation}
\sum_{m+a>0}(m+a)^0=\left[\left[\frac{1}{2}-a\right]\right],\quad
\sum_{m+a>0}(m+a)^1=-\frac{1}{2}\left[\left[\frac{1}{2}-a\right]\right]^2+\frac{1}{24},
\label{zeta}
\end{equation}
where $[[x]]$ denotes the element of ${\bf Z}+x$ with
the smallest absolute value.
The choice between $\pm1/2$ for $[[1/2]]$ does not affect the arguments below.
Using this formula, we obtain the vacuum energy of each twisted sector as
follows.
\begin{equation}
L_0=:L_0:-\frac{1}{2}+\left|\left[\left[\frac{k}{N}\right]\right]\right|.
\end{equation}
The left moving part of $U(1)_L$ charge $Q_{U(1)_L}$
is defined as one satisfying the following relations.
\begin{eqnarray}
&&[Q_{U(1)_L},\psi^A]=\psi^A,\quad
[Q_{U(1)_L},\ol\psi_A]=-\ol\psi_A,\nonumber\\{}
&&[Q_{U(1)_L},z^A]=z^A,\quad
[Q_{U(1)_L},\ol z_A]=-\ol z_A.
\end{eqnarray}
Such an operator is given by
\begin{equation}
Q_{U(1)_L}=
\sum_{m=-\infty}^\infty\sum_{A=1,2}\left(
           \frac{1}{-m+k/N}z^A_{m-\frac{k}{N}}\ol z_{A,-m+\frac{k}{N}}
           +\psi^A_{m+\frac{1}{2}-\frac{k}{N}}\ol\psi_{A,-m-\frac{1}{2}+\frac{k}{N}}
           \right).
\end{equation}
Regularizing this by (\ref{zeta}), we obtain
\begin{eqnarray}
Q_{U(1)_L}&=&:Q_{U(1)_L}:
+2\left[\left[\frac{k}{N}\right]\right]
+2\left[\left[\frac{1}{2}-\frac{k}{N}\right]\right]\nonumber\\
&=&\left\{\begin{array}{c}
:Q_{U(1)_L}:+1,\quad([[k/N]]>0),\\
:Q_{U(1)_L}:-1,\quad([[k/N]]<0).
\end{array}\right.
\end{eqnarray}
Namely, the ground state has $Q_{U(1)_L}=\pm1$ depending on the signature of $[[k/N]]$.
For the right moving part,
we can get the following result by
replacing $k$ in the result of the left moving part by $-k$:
\begin{equation}
\wt Q_{U(1)_L}=\left\{\begin{array}{c}
:\wt Q_{U(1)_L}:-1,\quad([[k/N]]>0),\\
:\wt Q_{U(1)_L}:+1,\quad([[k/N]]<0).
\end{array}\right.
\end{equation}
Therefore, total $U(1)_L$ charge $Q_{U(1)_L}+\wt Q_{U(1)_L}$
of the ground state is always zero.

Now, we should construct the Fock space by exciting the ground state with oscillators
and should impose GSO and $\Gamma$ projections.
The result for each case is as follows.
All results holds for A or B type theories.
\begin{description}
%%%%%%%%%%%%%%%%%%%%%%%%%%%%%
\item[type II on ${\bf R}^4/{\bf Z}_Q$]\mbox{}

The ground state of type II theories is GSO odd
and it should be excited by at least one fermionic oscillator.
If we restrict our attention to the case with $0<k/N<1/2$ for simplicity,
fermionic oscillators with the smallest energy are
$\ol\psi_{A,-1/2+k/N}$ in the left moving part and
$\wt\psi^A_{-1/2+k/N}$ in the right moving part.
By exciting the ground state with these oscillators, we have four massless states for each $k$.
\begin{equation}
\ol\psi_{A,-1/2+k/2}|0\rangle_L\otimes
\wt\psi^B_{-1/2+k/2}|0\rangle_R.
\end{equation}
Therefore, there is no tachyonic mode.
These massless modes correspond to blow up moduli parameters of the singularity.
%%%%%%%%%%%%%%%%%%%%%%%%%%%%%
\item[type 0 on ${\bf R}^4/{\bf Z}_Q$]\mbox{}

In the type 0 theory, in addition to the states obtained in type II case,
states with opposite fermion numbers are allowed.
Because we are focusing on tachyonic states now,
the only oscillators we can use are
$z^A_{-k/Q}$ in the left moving part and
$\ol{\wt z}_{A,-k/Q}$ in the right moving part. (We assume $0<k/N<1/2$.)
By exciting both vacuums in the left and right moving part by $P$ oscillators,
we obtain
\begin{equation}
z^{A_1}_{-k/Q}\cdots z^{A_P}_{-k/Q}|0\rangle_L\otimes
\ol{\wt z}_{B_1,-k/Q}\cdots\ol{\wt z}_{B_P,-k/Q}|0\rangle_R.
\label{tachyonicstate}
\end{equation}
These states belong to
\begin{equation}
{\bf(P+1)}\times{\bf(P+1)}={\bf(2P+1)}+{\bf(2P-1)}+\cdots+{\bf1},
\label{multi}
\end{equation}
representation of $SU(2)_R=SU(2)_D$ symmetry.
The mass of these states is
\begin{equation}
M^2=M_T^2+\frac{4}{l_s^2}\frac{k(P+1)}{Q},
\label{massE}
\end{equation}
where $M^2_T=-2/l_s^2$ is mass$^2$ of the bulk tachyon field.
For small $P$ (\ref{tachyonicstate}) gives tachyonic modes.
%%%%%%%%%%%%%%%%%%%%%%%%%%%%%
\item[type II on ${\bf R}^4/{\bf Z}^{(F_S)}_{2Q}$]\mbox{}

In type II theory on ${\bf R}^4/{\bf Z}^{(F_S)}_{2Q}$, we have the following spectrum.
For even $k$ we have the same spectrum with 
type II on ${\bf R}^4/{\bf Z}_{2Q}$ and we have no tachyonic modes.
For odd $k$, we should take account of the difference
between ${\bf Z}_{2Q}$ and ${\bf Z}^{(F_S)}_{2Q}$.
In GS formalism, the boundary condition for the fermion fields $S^\alpha$ is given as
\begin{equation}
S^\alpha(\sigma+2\pi)=(-)^kU(\gamma^k)S^\alpha(\sigma),
\end{equation}
where $U(\gamma^k)$ is a spinor representation of $\gamma^k\in SU(2)_L$.
The nontrivial monodromy causes the negative sign for odd $k$.
This change of the boundary condition for the GS fermions corresponds to
the reversal of the GSO projection in the NSR formalism.
Due to this change of the GSO projection,
we have the same tachyonic spectrum with
type 0 on ${\bf R}^4/{\bf Z}_{2Q}$.
Therefore, the tachyonic spectrum of this theory is obtained by replacing
$Q$ by $2Q$, and $k$ by $2k'$,where $k'\in{\bf Z}+1/2$
in (\ref{tachyonicstate}) and (\ref{massE}) as follows.
\begin{equation}
z^{A_1}_{-k'/Q}\cdots z^{A_P}_{-k'/Q}|0\rangle_L\otimes
\ol{\wt z}_{B_1,-k'/Q}\cdots\ol{\wt z}_{B_P,-k'/Q}|0\rangle_R.
\label{tachyonicstateO}
\end{equation}
\begin{equation}
M^2=M_T^2+\frac{4}{l_s^2}\frac{k'(P+1)}{Q},
\label{massO}
\end{equation}
The multiplicity of each level is given by (\ref{multi}).
%%%%%%%%%%%%%%%%%%%%%%%%%%%%%
\item[type 0 on ${\bf R}^4/{\bf Z}^{(F_R)}_{2Q}$]\mbox{}

On this orbifold, the boundary condition for the right moving NSR fermion fields
$\wt\psi^A$ is given as
\begin{equation}
\wt\psi^A(\sigma+2\pi)=-(-)^ke^{2\pi ik/2Q}\wt\psi^A(\sigma).
\end{equation}
Because of the extra factor $(-)^k$ we have no NS-NS sector for odd $k$.
For even $k$, we have the same spectrum with
type 0 on ${\bf R}^4/{\bf Z}_{2Q}$ before ${\bf Z}^{(F_R)}_{2Q}$ projection.
However, these tachyonic states
have $(-1)^{F_R}=-1$ and are projected out by ${\bf Z}^{(F_R)}_{2Q}$ projection.
Therefore, this configuration has no tachyonic spectrum.

\end{description}

%%%%%%%%%%%%%%%%%%%%%%%%%%%%%%%%%%%%%%%%%%%%%%%%%%%%%%%%%%%%%%%%%%%
\section{Gauge fields on NS5-branes}\label{RR.sec}
In the R-R sector, the zero point energy contributions from bosons and fermions always cancel
and ground states always give massless modes.
Therefore, tachyonic modes do not exist.
However, it is worth seeing how these twisted modes reproduce
gauge fields on coincident $Q$ NS5-branes.

Before GSO and $\Gamma$ projections, ground states of the R-R sector degenerate
and belong to $({\bf2}_L+{\bf2}_R)\times({\bf2}_L+{\bf2}_R)$ representation
of $SO(4)\subset P(1,5)$. (Now we are using light-cone formalism.)
After the projections,
we obtain the following massless spectrum in each case.
In what follows, we assume that ${\bf2}_L$ and ${\bf2}_R$ states have even and odd
worldsheet fermion numbers, respectively.
\begin{description}
%%%%%%%%%%%%%%%%%%%%%%%%%%%%%%%%%%%%%%%%
\item[type IIA on ${\bf R}^4/{\bf Z}_Q$]\mbox{}

In type IIA theory, GSO projection operator is $[(1-(-)^{F_L})/2][(1+(-)^{F_R})/2]$.
Therefore, we obtain massless states belonging to the vector representation
${\bf2}_R\times{\bf2}_L={\bf4}_{\rm vec}$ for each $k$.
On the other hand, a zero-mode of the R-R $3$-form field gives one six-dimensional $U(1)$
vector field.
Putting them together, we have $Q$ $U(1)$ gauge fields.
These correspond to Cartan part of $U(Q)$ gauge field on type IIB NS5-branes.
%%%%%%%%%%%%%%%%%%%%%%%%%%%%%%%%%%%%%%%%
\item[type IIB on ${\bf R}^4/{\bf Z}_Q$]\mbox{}

In type IIB theory, GSO projection operator is $[(1+(-)^{F_L})/2][(1+(-)^{F_R})/2]$.
Therefore, we obtain massless states belongs to ${\bf2}_L\times{\bf2}_L={\bf3}_L+{\bf1}$
for each $k$.
Furthermore, zero modes of the R-R $4$-form field and the R-R $2$-form field also belong
to ${\bf3}_L+{\bf1}$.
Putting them together,
we have $Q$ self-dual two-form gauge fields (${\bf3}_L$) and $Q$ scalar fields (${\bf1}$).
This corresponds to `Cartan part' of gauge fields on type IIA NS5-branes.
%%%%%%%%%%%%%%%%%%%%%%%%%%%%%%%%%%%%%%%%
\item[type IIA on ${\bf R}^4/{\bf Z}^{(F_S)}_{2Q}$]\mbox{}

In this case, we have the following spectrum.
For even $k$, we have the same spectrum with 
type IIA on ${\bf R}^4/{\bf Z}_{2Q}$.
Namely, we have one $U(1)$ vector fields for each $k$.
For odd $k$, due to the change of the GSO projection which we mentioned in the last section,
we have a mode in ${\bf2}_L\times{\bf2}_R$ rather than ${\bf2}_R\times{\bf2}_L$.
(The first and second factors represent left and right moving parts respectively.)
This gives a vector field again for each $k$.
Putting together them and a zero modes of R-R three form field,
we have $2Q$ $U(1)$ vector fields.
These correspond to Cartan part of $U(Q)\times U(Q)$ gauge fields on
type 0B NS5-branes.
%%%%%%%%%%%%%%%%%%%%%%%%%%%%%%%%%%%%%%%%
\item[type IIB on ${\bf R}^4/{\bf Z}^{(F_S)}_{2Q}$]\mbox{}

For even $k$, we have same spectrum with 
type IIB on ${\bf R}^4/{\bf Z}_{2Q}$.
Namely, we have one self-dual two-form gauge field and one scalar field
for each $k$.
For odd $k$, due to the opposite GSO projection,
we obtain fields with the opposite chirality belonging to
${\bf2}_R\times{\bf2}_R={\bf3}_R+{\bf1}$.
These are an anti-self-dual two-form field (${\bf3}_R$) and a scalar field (${\bf1}$).
Putting together them and zero modes of R-R self-dual four-form field and
R-R two-form field, we have $Q$ unconstrained two-form fields and
$2Q$ scalar fields.
These correspond to the `Cartan part' of massless gauge fields on
type 0A NS5-branes.
%%%%%%%%%%%%%%%%%%%%%%%%%%%%%%%%%%%%%%%%
\item[type 0A on ${\bf R}^4/{\bf Z}_Q$]\mbox{}

In type 0A theory, the GSO projection operator is $(1-(-)^{F_L+F_R})/2$.
Therefore, we obtain massless states belonging to
${\bf2}_R\times{\bf2}_L+{\bf2}_L\times{\bf2}_R={\bf4}_{\rm vec}+{\bf4}_{\rm vec}$
for each $k$.
Putting together them and zero modes of two R-R $3$-form fields, we have $2Q$ $U(1)$ gauge fields.
This corresponds to the Cartan part of $U(Q)\times U(Q)$ gauge fields on type 0B NS5-branes.
%%%%%%%%%%%%%%%%%%%%%%%%%%%%%%%%%%%%%%%%
\item[type 0B on ${\bf R}^4/{\bf Z}_Q$]\mbox{}

In type 0B theory, the GSO projection operator is $(1+(-)^{F_L+F_R})/2$.
Therefore, we obtain massless states belonging to
 ${\bf2}_L\times{\bf2}_L+{\bf2}_R\times{\bf2}_R={\bf3}_L+{\bf3}_R+{\bf1}+{\bf1}$
for each $k$.
Putting together them and zero modes of the unconstrained R-R $4$-form field
and two R-R $2$-form fields,
we have $Q$ unconstrained two-form gauge fields (${\bf3}_L+{\bf3}_R$)
and $2Q$ scalar fields (${\bf1}$).
This corresponds to the `Cartan part' of the gauge fields on type 0A NS5-branes.
%%%%%%%%%%%%%%%%%%%%%%%%%%%%%%%%%%%%%%%%
\item[type 0A on ${\bf R}^4/{\bf Z}^{(F_R)}_{2Q}$]\mbox{}

Due to the change of boundary condition of the left moving fermions
by $(-1)^{F_R}$ monodromy, there is no R-R sector for odd $k$.
For even $k$, we have two $U(1)$ vector fields from each $k$ like
type 0A on ${\bf R}^4/{\bf Z}_{2Q}$.
However, one of them are projected out by the ${\bf Z}^{(F_R)}_{2Q}$ projection
because they have opposite $(-)^{F_R}$ quantum numbers.
Similarly, one of zero modes of two R-R three-form fields is projected out.
Putting them together, we have $Q$ $U(1)$ vector fields.
These correspond to the Cartan part of $U(Q)$ gauge fields on
type IIB NS5-branes.
%%%%%%%%%%%%%%%%%%%%%%%%%%%%%%%%%%%%%%%%
\item[type 0B on ${\bf R}^4/{\bf Z}^{(F_R)}_{2Q}$]\mbox{}

We have no R-R sector for odd $k$.
For even $k$, we have same spectrum with 
type 0B on ${\bf R}^4/{\bf Z}_{2Q}$ before the ${\bf Z}^{(F_R)}_{2Q}$ projection.
Namely, we have one unconstrained two-form gauge fields and two scalar fields
for each $k$.
In addition to them, we have one unconstrained two-form field as zero mode of unconstrained
R-R four-form field and two scalar modes as zero modes of two R-R two-form fields.
However, by the ${\bf Z}^{(F_R)}_{2Q}$ projection, half of them are projected out.
As a result, we have $Q$ self-dual two-form fields and $Q$ scalar fields.
These correspond to the `Cartan part' of the massless gauge fields on
type IIA NS5-branes.
\end{description}

%%%%%%%%%%%%%%%%%%%%%%%%%%%%%%%%%%%%%%%%%%%%%%%%%%%%%%%%%%%%%%%%%%%
\section{Tachyon modes on NS5-brane solution}\label{mode.sec}
In Section \ref{NSNS.sec}, we showed that tachyonic twisted modes arose in type 0
on ${\bf R}^4/{\bf Z}_Q$ and type II on ${\bf R}^4/{\bf Z}^{(F_S)}_{2Q}$,
both which are dual to type 0 NS5-branes.
What are counterparts for these tachyonic modes on the NS5-branes side?
We show that these can be reproduced as modes of the bulk tachyon fields
localized on a NS5-brane classical solution.

In order to analyze modes, we need a classical solution of supergravity.
The relevant part of supergravity action is
\begin{equation}
S=\frac{1}{(2\pi)^7l_s^8}\int d^{10}x\sqrt{-g}\frac{1}{e^{2\phi}}
    \left(R+4(\partial\phi)^2-\frac{(2\pi l_s)^4}{2\cdot3!(2\pi)^2}H_3^2\right).
\end{equation}
This is common to type IIA, IIB, 0A and 0B supergravities.
We use the convention in which the gauge flux is quantized as follows.
\begin{equation}
\oint_{{\bf S}^3}H_3=2\pi Q.
\end{equation}
The classical solution representing parallel NS5-branes is\cite{hetero,dufflu}
\begin{equation}
ds^2=\eta_{\mu\nu}dx^\mu dx^\nu+f(x^i)(dx^i)^2,\quad
e^{2\phi}=f(x^i).
\end{equation}
where $x^\mu$ ($\mu=0,\ldots,5$) and $x^i$ ($i=6,\ldots,9$) are
coordinates along parallel and transverse directions, respectively.
The function $f(x^i)$ is a harmonic function on the $x^i$ plane,
which is related to a fivebrane density $\rho_{\rm NS5}(x^i)$
by the following Laplace equation.
\begin{equation}
\sum_{i=6}^9\left(\frac{\partial}{\partial x^i}\right)^2f(x^i)=-(2\pi l_s)^2\rho_{\rm NS5}(x^i).
\end{equation}
A solution for $Q$ coincident NS5-branes located at a point $x^i=0$
is obtained by setting
\begin{equation}
\rho_{\rm NS5}(x^i)=Q\sum_{n=-\infty}^\infty\delta(x^6,x^7,x^8,x^9-2\pi nR),
\label{localcharge}
\end{equation}
where we took the compactification of the $x^9$ direction into account.
In this case, $f(x^i)$ is given as
\begin{equation}
f(x^i)=1+\sum_{n=-\infty}^\infty\frac{r_0^2}{(x^i)^2+(x^9-2\pi nR)^2},\quad
r_0^2=l_s^2Q.
\end{equation}

Tachyonic modes obtained in Section \ref{NSNS.sec}
belong to ${\bf(P+1)}\times{\bf(P+1)}$ representation of $SU(2)_D=SU(2)_L$ rotation symmetry.
If this is $({\bf P+1},{\bf P+1})$ representation of $SU(2)_1\times SU(2)_2$,
the structure of the spectrum is similar to that of modes of a scalar field
in $SO(4)=SU(2)_1\times SU(2)_2$ symmetric potential.
This seems to imply that these modes are localized in the near horizon region $|x^i|\ll R$
where the effect of compactification can be neglected.
However, as we will show below, this is not true.
This degeneracy of spectrum is accidental one rather than result of some symmetry.

At first, let us consider the region
where $r=(\sum_{i=6}^9x_i^2)^{1/2}$ satisfies
\begin{equation}
r\ll r_0\mbox{ and }r\ll R.
\label{and}
\end{equation}
In this region, the metric reduce to
\begin{equation}
ds^2=\eta_{\mu\nu}dx^\mu dx^\nu+r_0^2\frac{dr^2}{r^2}+r_0^2d\Omega_3^2.
\end{equation}
This metric has structure ${\bf R}^6\times{\bf R}^+\times{\bf S}^3$.
Unfortunately, the infinitely long throat structure of this solution implies
tachyon modes cannot be quantized.
Actually, using the action of tachyon field
\begin{equation}
S=-\frac{1}{(2\pi)^7l_s^8}\int d^{10}x\sqrt{-g}\frac{1}{2e^{2\phi}}\left[(\partial T)^2+M_T^2T^2\right],
\label{Taction}
\end{equation}
we obtain the following equation of motion.
\begin{equation}
\Delta_6T=\left[M_T^2-\frac{1}{r_0^2r}\partial_rr^3\partial_r
                        -\frac{1}{r_0^2}\Delta_{{\bf S}^3}\right]T,
\end{equation}
where $\Delta_{\bf S}^3$ is Laplacian on unit ${\bf S}^3$ and $\Delta_6$ is Laplacian
along $x^\mu$, whose eigenvalue gives the mass$^2$ of the modes on the fivebrane.
The eigenfunction of the operator in the second term in the bracket in the right hand side is
$r^n$ and its eigenvalue is $s=n(n+2)/r_0^2$.
If $s<1/r_0^2$, the wave function diverges at $r\rightarrow0$ or $r\rightarrow\infty$.
They represent modes falling into horizon and modes living outside the throat.
If $s>1/r_0^2$,
the wave function can propagate along the throat.
The existence of these mode are connected with
the non-vanishing of Hawking radiation and
that of absorption cross section for bulk field\cite{MaSt,MS}.
This result implies that tachyon modes cannot be trapped in the near horizon region.
So, we need to consider modes outside it.
If $r_0\ll R$, the outside of the region (\ref{and}) is flat
and modes cannot be trapped.
Therefore we assume
\begin{equation}
R\ll r_0.
\label{Rr0}
\end{equation}
In this case, the region with $r\gg R$ is described by a smeared NS5-brane solution.
A smeared NS5-brane solution with charge $Q$ distributed around the circle uniformly
is specified by the following NS5-brane density.
\begin{equation}
\rho_{\rm NS5}(x^i)=\frac{Q}{2\pi R}\delta^3(x^6,x^7,x^8).
\end{equation}
The harmonic function $f(x^i)$ is given by
\begin{equation}
f(x^i)=1+\frac{r_0'}{r},\quad
r_0'=\frac{l_s^2Q}{2R}.
\label{smeared}
\end{equation}
Now, $r$ is defined by $r=\sum_{i=6,7,8}(x^i)^2$.
The near horizon ($r\ll r_0'$) metric of this solution is
\begin{equation}
ds^2=\eta_{\mu\nu}dx^\mu dx^\nu+\frac{r_0'}{r}(dz^2+dr^2)+r_0'rd\Omega_2^2.
\label{NHsmeared}
\end{equation}
We should notice that (\ref{Rr0}) is equivalent to $r_0\ll r_0'$
and the region described by (\ref{NHsmeared}) always exists.

On this manifold, we have the following equation of motion of tachyon.
\begin{equation}
\Delta_6T
=\left[M_T^2-\frac{r}{r_0'}\partial_9^2
           -\frac{1}{r_0'r}\partial_rr^2\partial_r
           -\frac{1}{r_0'r}\Delta_{{\bf S}^2}\right]T.
\end{equation}
This differential equation has the same structure with the Shr\"odinger equation for
a particle in a Coulomb potential.
Let us decompose $T$ as
\begin{equation}
T=\phi(r)Y_{L,m}(\theta_1,\theta_2)e^{-ip_9x^9}e^{-ip_\mu x^\mu},
\end{equation}
where $Y_{L,m}(\theta_1,\theta_2)$ is the spherical harmonic function on ${\bf S}^2$
and $p_9$ and $p_\mu$ are momenta along $x^9$ and $x^\mu$, respectively.
The mass on the fivebrane $M$ is given by $M^2=-\eta^{\mu\nu}p_\mu p_\nu$.
Then, the equation reduced to
\begin{equation}
M^2\phi=\left[M_T^2+\frac{r}{r_0'}p_9^2
           -\frac{1}{r_0'r}\partial_rr^2\partial_r
           +\frac{1}{r_0'r}L(L+1)\right]\phi.
\label{msqdiff}
\end{equation}
The Kaluza-Klein momenta $p_9$ is quantized as
\begin{equation}
p_9=\frac{m}{R},
\end{equation}
where $m$ is an integer in the case of ${\bf S}^1$ compactification
and is a half odd integer in the case of ${\bf S}^1_{(F_R)}$ compactification
because the tachyon field has $(-1)^{F_R}=-1$.
Via T-duality, $m$ is transformed into wrapping number,
and is identified with $k\in{\bf Z}$ in (\ref{tachyonicstate}) and (\ref{massE}) or
$k'\in{\bf Z}+1/2$ in (\ref{tachyonicstateO}) and (\ref{massO}).
One might be afraid that (\ref{msqdiff}) cannot reproduce the twisted mode spectrum
because the Kaluza-Klein momentum depends
on the compactification radius $R$ while the dependence is absent on the
twisted mode side.
This, however, is not the case.
By rescaling of radial coordinate
\begin{equation}
\rho=2p_9r,
\label{rhois}
\end{equation}
we obtain the $R$-independent expression
\begin{equation}
\frac{1}{\rho^2}\partial_\rho\rho^2\partial_\rho\phi=
\left[\frac{1}{4}-\frac{\lambda}{\rho}+\frac{L(L+1)}{\rho^2}\right]\phi,
\label{diffeq}
\end{equation}
where we defined $\lambda$ as follows.
\begin{equation}
\lambda=\frac{r_0'}{2p_9}(M^2-M_T^2).
\label{lambdais}
\end{equation}
This number corresponds to the principal quantum number of a particle
in the Coulomb potential
and is quantized as we show below.

For the tachyon field not to diverge at the origin and infinity,
the function $\phi(\rho)$ should behave like the following way
asymptotically.
\begin{equation}
\phi(\rho)\sim\rho^L\quad(\rho\rightarrow0),\quad
\phi(\rho)\sim e^{-\rho/2}\quad(\rho\rightarrow\infty).
\label{behavior}
\end{equation}
Let us expand the function $\phi(\rho)$ as follows.
\begin{equation}
\phi(\rho)=\sum_na_n\rho^ne^{-\rho/2}.
\label{expandf}
\end{equation}
Then, the differential equation (\ref{diffeq}) gives the following
relation among the coefficients $a_n$.
\begin{equation}
\left[n(n+1)-L(L+1)\right]a_n=(n-\lambda)a_{n-1}
\end{equation}
If we put $n=L$, this equation gives $a_{L-1}=0$
and it is consistent with the $\rho\rightarrow0$ behavior of $\phi(\rho)$.
To reproduce $r\rightarrow\infty$ behavior in (\ref{behavior}),
only finite number of coefficients $a_n$ can be non-zero.
For this condition to be satisfied, the following values are allowed for $\lambda$.
\begin{equation}
\lambda=L+1,L+2,L+3,\ldots.
\end{equation}
As a result, we obtain the mass of the mode.
\begin{equation}
M^2=M_T^2+\frac{4m}{l_s^2Q}\lambda
\label{m6is}
\end{equation}
If we identify $\lambda$ with $P+1$,
this result exactly reproduces the tachyonic spectrum of the twisted modes
(\ref{massE}) and (\ref{massO})!

For fixed $\lambda(=P+1)$, all states with $L$ smaller than $\lambda$ are degenerate.
Therefore, multiplicity is
\begin{equation}
{\bf(2P+1)}+{\bf(2P-1)}+\cdots+{\bf1}={\bf(P+1)}\times{\bf(P+1)}.
\end{equation}
This is completely same with the multiplicity (\ref{multi}) of the twisted modes.

%%%%%%%%%%%%%%%%%%%%%%%%%%%%%%%%%%%%%%%%%%%%%%%%%%%%%%%%%%%%%%%%%
\section{Discussion}
In this paper, we restricted our attention to the tachyonic modes
by the following reason.
\begin{itemize}
\item The structure of spectrum of twisted modes is very simple.
Because many oscillators contribute to massive states,
the massive spectrum is more complicated.
\item On classical solution side, the identification of modes is very easy.
Tachyonic modes can come from only the tachyon field.
In order to identify massive twisted modes with modes of bulk fields,
we need to analyze their quantum numbers in more detail.
\item Because the tachyon field is a scalar,
it is very easy to calculate the eigen mode on the classical solution.
\end{itemize}
Of course, it is an interesting problem to investigate to what extent this correspondence
holds.
Especially, it is important to understand how the gauge fields on NS5-branes
are described in the context of supergravity.
However, leaving technical obstacles, it seems difficult
to reproduce all gauge fields obtained as twisted modes.
According to the argument in the AdS/CFT correspondence,
supergravity modes corresponds only to gauge invariant operator.
In fact, in \cite{CHS}, only the gauge singlet part of gauge fields are
obtained as modes of bulk R-R fields.
Therefore, classical modes on NS5-brane solutions may be matched off against
gauge singlet twisted modes at orbifold singularity.

In Section \ref{mode.sec}, we used the smeared NS5-brane metric (\ref{NHsmeared})
to obtain the discrete tachyonic modes.
The metric (\ref{NHsmeared}) is available only in the region
\begin{equation}
R\ll r\ll r_0'.
\label{rrr}
\end{equation}
Therefore, analysis of tachyon modes is valid
only when the support of the wave function of the mode is in this region.
In the rest of this section, let us discuss this condition.
Using (\ref{rhois}), the upper bound in (\ref{rrr}) is rewritten as
\begin{equation}
\rho\ll\frac{l_s^2Qm}{R^2}.
\end{equation}
It is known that the following equation holds for a solution of (\ref{diffeq})
\begin{equation}
\ol{\rho^{-1}}\equiv\frac{\int_0^\infty(1/\rho)\rho^2\phi^2(\rho)d\rho}{\int_0^\infty \rho^2\phi^2(\rho)d\rho}
=\frac{1}{2\lambda}.
\end{equation}
In the context of the quantum mechanics on the Coulomb potential,
this implies that expectation value of the potential energy is proportional to the total
energy.
So, we can use $P\sim\lambda$ as a typical value of $\rho$.
Then we obtain the following bound.
\begin{equation}
P\ll\frac{l_s^2Qm}{R^2}.
\label{limit2}
\end{equation}
On the other hand,
on the Kaluza-Klein monopole side,
we neglect the first term in (\ref{vis}) to obtain the orbifold metric (\ref{flat}).
This is possible for $r^2\sim\wt RQ|x^i|\ll\wt R^2Q^2$.
The value $r^2=|z^A|^2$ can be estimated as an expectation value of
the following operator.
\begin{eqnarray}
&&\frac{1}{2\pi}\int_0^{2\pi}:z^A(\sigma)\ol z_A(\sigma):d\sigma\nonumber\\
&&\hspace{1em}=l_s^2\sum_{m=-\infty}^\infty\sum_{A=1,2}
\left(\frac{:z^A_{m-k/N}\ol z_{A,-m+k/N}:}{(m-k/N)^2}
     +\frac{:\wt z^A_{m+k/N}\ol{\wt z}_{A,-m-k/N}:}{(m+k/N)^2}\right).
\end{eqnarray}
On the states (\ref{tachyonicstate}), this gives
\begin{equation}
r^2\sim l_s^2\frac{QP}{k}.
\end{equation}
Therefore, the applicable limit is
\begin{equation}
\frac{P}{Qk}\ll\frac{l_s^2}{R^2}.
\label{limit}
\end{equation}
This is the same as (\ref{limit2}).

Finally, let us discuss the lower bound in (\ref{rrr}).
This condition demand the tachyon modes not to see the localization of NS5-branes.
If we use $P$ as the typical value of $\rho$ again, the lower bound in (\ref{rrr}) is
rewritten as
\begin{equation}
\frac{k}{P}\ll1.
\label{KP}
\end{equation}
In \cite{Hmonopole}, the relation between a Kaluza-Klein monopole and
a localized NS5-brane are argued and it is shown that
zero modes of NS-NS two-form field on the Kaluza-Klein monopoles,
which are also regarded as twisted modes at the singularity,
play an important role.
However, in our analysis, we did not take account of the interactions among twisted sectors.
Therefore, it is valid only when the interaction can be neglected.
This argument seems consistent with condition (\ref{KP})
because, roughly speaking, the interaction between the $B$ field and strings is proportional to
the winding number $k$ and
the probability that a string exist near the singularity becomes smaller
when $P$ becomes larger.
%%%%%%%%%%%%%%%%%%%%%%%%%%%%%%%%%%%%%%%%%%%%%%%%%%%%%%%%
\section*{Acknowledgement}
I would like to thank B. Zwiebach
for careful reading of the manuscript.
I also appreciate hospitality of the organizers of
Summer Institute '99 at Yamanashi, Japan,
where a part of this work was discussed. 

%%%%%%%%%%%%%%%%%%%%%%%%%%%%%%%%%%%%%%%%%%%%%%%%%%%%%%%%

\end{document}